# Pressure effect on multiferroic CuBr$_2$

Li Zhao[1], Ching-Chien Li[1], Chun-Chuen, Yang[2] and Maw-Kuen Wu[1]

[1]Institute of Physics, Academia Sinica, Taipei 11529, Taiwan

[2]Department of Physics, Chung Yuan Christian University, Taoyuan 32023, Taiwan

Abstract

The quasi-1D spin chain compound CuBr$_2$ has been found to be multiferroic below $T_N$ (~73.5K) under ambient pressure, in which the spontaneous electric polarization is induced by emerging spin spiral ordering propagating along *b*-axis. Herein we studied the hydrostatic pressure effect on the magnetic, dielectric and structural properties of CuBr$_2$. The multiferroic transition temperature is greatly enhanced under hydrostatic pressure. From ambient to about 1 GPa (the limit of our homemade apparatus), $T_N$ increases unprecedentedly by more than 20K, and no sign of saturation is observed in our experiments. Meanwhile the corresponding dielectric loss keeps rather low (<<0.1). Further synchrotron-based high pressure X-ray diffraction measurements reveal that there is no pressure-induced structural phase transition in CuBr$_2$ up to 10.2 GPa. Upon increasing pressure, the *b*-axis (along the spin chain) just changes slowly while transverse *a*- and *c*- axis parameters shrink much more greatly. Pressure greatly reduces the separation between spin chains and enhance the inter-chain coupling interactions in CuBr$_2$, which results in the giant increase of multiferroic critical temperate. Our finding suggests a new effective way to improve the known multiferroic systems towards practical high temperature multiferroicity.

## 1. Introduction

The revival of magnetoelectric research [1] at the beginning of this millennium has been ignited by the discovery of the multiferroicity in some magnetically frustrated manganites.[2,3]. In these multiferroic materials, the spin-driven ferroelectricity and concomitant large magnetoelectric coupling were discovered. The spontaneous electric polarization found in these frustrated manganites is of magnetic origin, i.e., induced by complex spin structure which break the inversion symmetry. The electric polarization can be inversed by changing the magnetic state via applying external magnetic field, and vice versa [4]. The studies on these fascinating multiferroics materials open new opportunities not only for fundamental physics of strongly correlated materials, but also for potential application of highly efficient multi-functional devices as sensors and memory devices.[5].

Most known spin-driven ferroelectricity has also been found existing in many helimagnetic oxides, including manganites[2,3], cuprates[6-8], ferrates [99] and nickelates[11]. A well-accepted key microscopic mechanism for their multiferroicity comes down to the inverse Dzyaloshinskii-Moriya(DM) interaction, which is an antisymmetric relativistic correction to the superexchange coupling. It can also be expressed in an equivalent spin current picture by Katsura, Nagaosa and Balatsky (KNB model). The microscopic polarization $P_{ij}$ induced by neighboring spins can be formulated as $P_{ij} = A\hat{e}_{ij} \times (S_i \times S_j)$ where the coupling coefficient $A$ is determined by the spin-orbit coupling and exchange interactions, and $\hat{e}_{ij}$ is the unit vector connecting the sites *i* and *j*. This has worked well for many helix magnetic materials [12].

In spite of close tie between the polarization and magnetic ordering, the transition temperature of most known multiferroic materials is far below room temperature (usually below 40 K), since that their spin-ordered phases comes from highly frustrated competing magnetic interactions with strong fluctuations and therefore multiferroicity usually exists only in rather narrow ranges at low temperature [13]. The application of multiferroicity has been mostly hampered by their low transition temperature. Till now, only a few compounds have been found with relatively high critical temperatures. The most famous example is copper (II) oxide, CuO. The ferroelectricity is induced by its incommensurate spiral magnetic structure (AF2 phase) existing between 213 K and 230 K [6]. In some hexaferrites, multiferroicity even above room temperature has been reported [14]. But the high dielectric loss (tan δ > 1) exists due to their semiconducting nature, preventing accurate dielectric measurements and potential device applications in the future. Until now, the reproducibility of single-phase room temperature multiferroics at ambient still seems elusive as room-temperature superconductors. At present, there is still a long way to go in developing applicable multiferroic materials with both high critical temperature and low electric loss.

Pressure as a clean, continuous and systematic tuning parameter, has been widely used in the field of strong correlated electron systems. Recent theoretical calculation has suggested that applying high pressure (around 20 - 40 GPa) can induce a giant stabilization of the aforementioned multiferroic AF2 phase of CuO even above 300K with large polarization [15]. But till now, this predicted room temperature multiferroicity of CuO under high pressures have not yet to be examined. There have been some studies on hydrostatic pressure effects on some known multiferroic oxide systems, as rare-earth manganites [16], ferrites [17] etc. Some pressure-induced multiferroic phases or enhanced ferroelectric polarization have been observed. But the corresponding critical temperatures are still rather low and the corresponding pressure dependence is rather weak in these oxide systems, partially due to the rigid oxide lattice and low compressibility.

Recently, people discovered some new non-oxide multiferroic compounds as cupric halogenides and oxyhalides[18-20]. Among these, $CuBr_2$ has attracted particular attention. As the isostructural $CuCl_2$ [18], $CuBr_2$ is a chemically simple quasi-1D antiferromagnetic spin 1/2 quantum chain system. As shown in Fig. 1(a), its spin chains are made of $CuBr_4$ square units via edge-sharing running along b-axis, which resembles the $CuO_4$-square-based spin-chain multiferroic cuprates as $LiCu_2O_2$ [7] and $LiCuVO_4$ [8, 9]. A spiral magnetic structure (as shown in Fig. 1(b)) emerges below Néel temperature ($T_N$ ~ 73.5K) due to competing nearest-neighbor (NN) ferromagnetic and next-nearest-neighbor (NNN) antiferromagnetic exchange interactions along the chain, which induced ferroelectricity via the aforementioned inverse Dzyaloshinskii-Moriya mechanism. The most remarkable feature of $CuBr_2$ is its relative high $T_N$ (close to liquid nitrogen temperature, much higher than its isostructural cousin $CuCl_2$ (24K) [18] and other multiferroic cuprate compounds (23K for $LiCu_2O_2$ [21,18] and 2.4K for $LiCuVO_4$ [9] ). The abnormally high $T_N$ in $CuBr_2$ remains still elusive although there have already been some theoretical explorations [2221, 23].

Herein, we systematically studied the hydrostatic pressure effect on multiferroic properties (magnetic susceptibility and dielectric constant) of $CuBr_2$ using our home-made high pressure cell, as well as characterizing the structure evolution using a diamond anvil cell (DAC). The great enhancement of multiferroic transition in $CuBr_2$ by high pressure is observed, which open up new potential towards room temperature multiferroic device application.

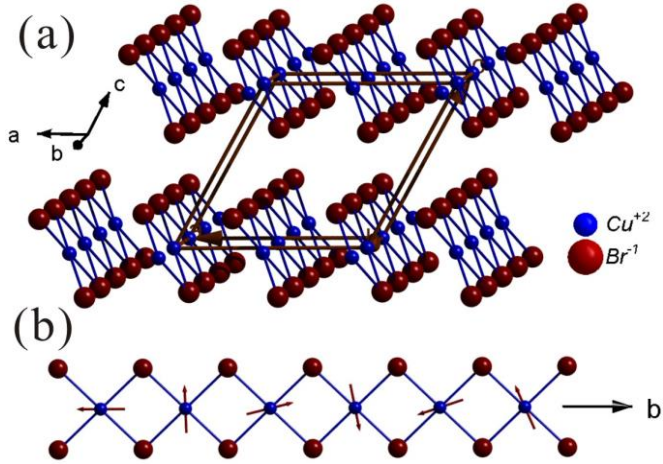

Figure 1. (a) Crystal structure of $CuBr_2$. (b) The spin chain consists of edge-sharing $CuBr_2$ plaquettes running along *b*-axis in the $CuBr_2$, with the arrows denote the cycloidal ordering of $Cu^{2+}$ s=1/2 spins below $T_N$.

## 2. Experimental

The sample preparation has been reported in our previous work [19]. For magnetic susceptibility measurement, About 20 mg $CuBr_2$ powder is loaded in a Teflon capsule, together with a high pure lead grain used as the monometer. Flourinert 77 was employed as a pressure-transmitting medium. The Teflon capsule is loaded in a piston-cylinder-type pressure cell, which is home-made of high-strength Beryllium Copper alloy and zirconia. The *in situ* pressure is determined by measuring Meissner effect of Pb under a magnetic field of 10 Oe [24] using a commercial superconducting quantum interference device (SQUID) magnetometer (MPMS-XL-5T, Quantum Design Co. Ltd.).

The temperature dependent magnetic susceptibility, $\chi(T)$, is measured in the field of 2000 Oe. Because of the antiferromagnetic nature of $CuBr_2$ and the limited volume of sample holder (the I.D. of Teflon capsule is just 1 mm), the magnetic response from our sample is rather weak compared with the background signal from the whole pressure cell. In order to obtain precise $\chi$, the SQUID raw response is collected with and without a sample and the difference signal is used to calculate the moment with a external program according to the tech note (No. 1014-213) from the Quantum Design Inc.

To study the dielectric properties of $CuBr_2$, we apply the silver paint was both sides of a thin pellet (0.2 mm thick) as electrodes to form a parallel plate capacitor, whose capacitance is proportional to the dielectric constant ($\varepsilon$). This sample is loaded in a home made Cu-Be pressure cell of piston-cylinder-type. Then the pressure cell is mounted on a homemade insertion compatible with a Quantum Design PPMS system. The mixture of Fluorinert 70 and 77 is used as a pressure-transmitting medium and a piece of Pb wire is also loaded to calibrate the pressure whose superconducting transition is measured with a lock-in amplifier (SR-830, Stanford Research System Inc.). The capacitance is measured with a high-precision LCR meter (Agilent E4980A, Keysight Inc).

In the present experimental setup, the systematic errors (mainly from residual impedance in the whole circuit) is hard to be simply compensated. The pressure dependence of the medium surrounding the sample is unknown at low temperature, as well as is the stray capacitance from the crowded wiring in the Teflon cell (6 enameled wires are used, with other 4 spared wires), But all these contributions do not shows any anomaly in the temperature range we concern. Furthermore, the soft $CuBr_2$ sample has much larger compressibility than oxide samples (as shown in our later structural characterization) and the decrease of the sample dimensions by the applied pressure cannot be negligible. Therefore, the accurate $\varepsilon$ can not be accessible in our experiments, and we just present the raw capacitance data (as shown in Fig. 3) other than $\varepsilon$.

In our experiments, the sweep rate is kept below 0.5 K/min to eliminate the temperature inhomogeneity of the high pressure cell. The difference of the results measured during the warming and cooling processes can be negligible, as exemplified the capacitance data for P = 0.53 GPa case shown in the Fig. 3(b).

To study the evolution of the lattice parameters under the hydrostatic pressure, and possible pressured-induced structural transition, the *in situ* high-pressure powder x-ray diffraction (HP-XRD) was measured at the beamline BL12B2 (owned by National Synchrotron Radiation Research Center) in the SPring-8 synchrotron in Japan, using a diamond anvil cell (DAC). All data are collected in room temperature. The Rietveld refinements of the diffraction patterns were performed using GSAS software package [25].

## 3. Results and Discussion

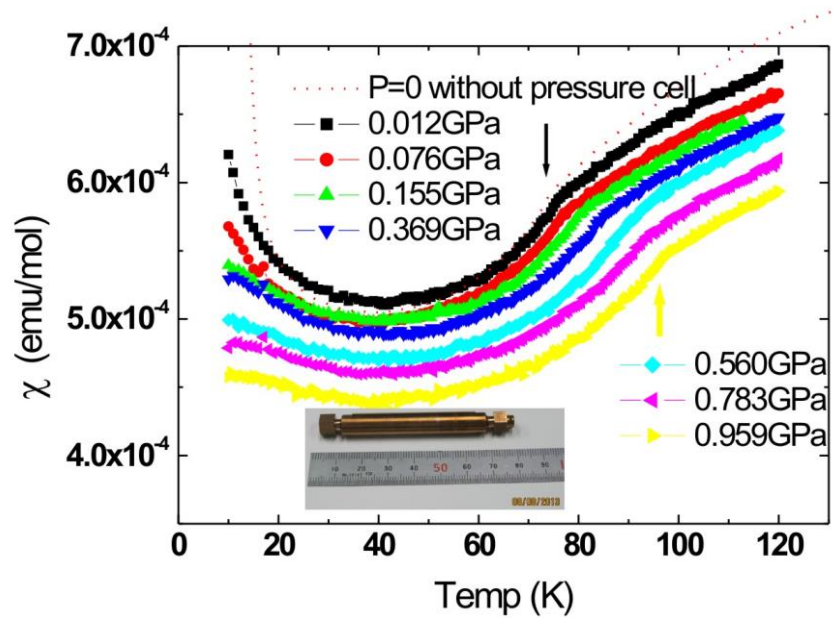

Fig. 2 Temperature dependent magnetic susceptibility of $CuBr_2$ measured under different hydrostatic pressure (0.012 to 0.959 GPa) using our home made pressure cell (shown in the inset). For clarity, all the curves are shifted vertically except for 0.012 GPa. The P = 0 data are also plotted (the dotted line) for comparison, measured without the pressure cell.

Fig. 2 shows the temperature dependence of magnetic susceptibility of $CuBr_2$ measured in H = 2000 Oe under different pressure. There is no discernable hysteresis between measurements during zero-field-cooling and field-cooling. For comparison, the $CuBr_2$ powder of around 50 mg are measured in a convention way (using a drinking straw and gelatin capsule) as the P = 0 case (the dotted line in Fig. 2). The small difference between this P = 0 curve and the P → 0 (0.012 GPa) one measured with the pressure cell are mostly attributed to imperfect subtraction of the background signal from the pressure cell, which can be negligible qualitatively. Above $T_N$, a broad hump has been reported previously at ambient [19]. A kink-like anomaly (marked by the arrows) occurs for all the cases, which denote the onset of a long-range magnetic ordering. This magnetic transition at ambient has been confirmed to ben a spiral-type one in our previous neutron diffraction experiment [19].

With increasing pressure, the kink-like anomaly shifts to higher temperatures gradually, while the up-turns at lower temperature are slightly suppressed accordingly. There is no other anomalous features emerging from ambient to about 1 GPa in the temperate range we concern. The main effect of external pressure is just push to the magnetic transition to higher temperature. At about 0.96GPa, the $T_N$ reaches to 95 K, more than 20 K higher than that at ambient.

At ambient, we have confirmed the ferroelectric nature of the transition at $T_N$ by corresponding dielectric measurements [19]. Herein, we carried out systematical dielectric measurements on the $CuBr_2$ sample in the pressure cell under a series of pressures. Fig. 3(a) shows the results of one typical run in zero field for the P = 0.55 GPa, measured during the warming process. As mentioned before, it is difficult to extract the intrinsic ε(T) of $CuBr_2$ accurately, and only raw capacitance data are presented. The most remarkable anomaly is the sharp rising of capacitance just below ~85K, which coincides well with the $T_N$ value acquired by the above magnetic susceptibility measurements (see the $T_N$ vs. T summarized in Fig. 5). This dielectric anomaly can be observed for all the testing frequencies ( 1 - 50 kHz), The frequency independence excludes the possible extrinsic factors such as the trapped interfacial charge carriers or circuit stray impedance, and suggests the intrinsic electric transition of $CuBr_2$ at $T_N$.

Furthermore, the corresponding dielectric loss shows frequency-independent anomaly at $T_N$ too, confirming a concomitant a ferroelectric nature of the transition at $T_N$. At present, we cannot measure the electric polarization directly due the limit of the experimental setup. It is noticeable that at high pressure, the dielectric loss is still quite small (<<0.1) the same as the ambient one. $CuBr_2$ keeps highly insulating with very low dielectric loss under high pressure, which surpasses most known high temperature multiferroic compounds as ferrites and CuO. Low loss of $CuBr_2$ makes it potentially appropriate for future device applications.

In Fig. 3(b), we show the temperature dependent capacitance under different pressures, measured at 30 kHz (which achieves the best SNR for all cases in our homemade apparatus). The variation of absolute capacitance values comes mostly from deformation of Teflon capsule and relative displacement of sample/monomer/wiring inside the cell caused by loading and releasing processes before each measurement. Due to the limited dimensions (to match the PPMS chamber of about 1 inch in diameter) and the large inside hole (I.D.=6.3mm, accommodating the capacitance Teflon capsule), the achievable highest pressure is less than 1 GPa .

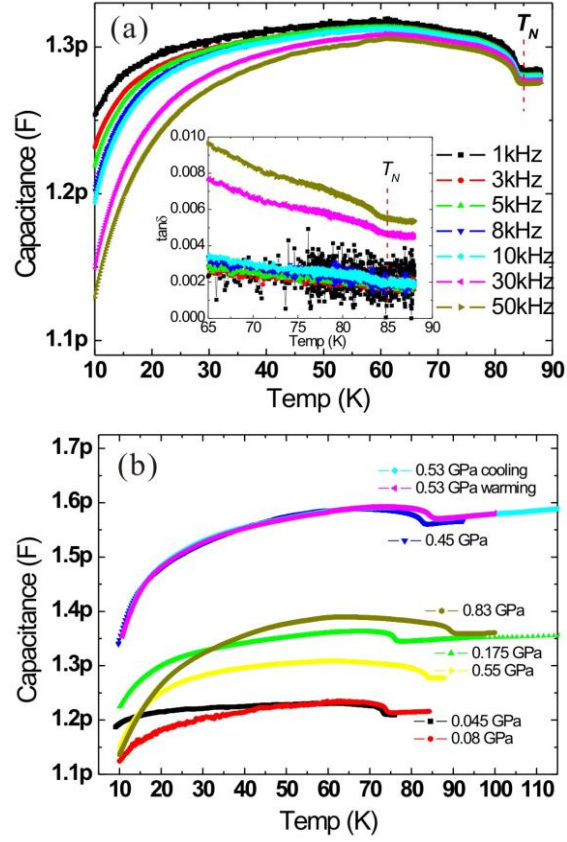

Fig. 3. (a) Temperature dependent capacitance of a CuBr2 sample under P=0.55GPa, measured at multiple frequencies (1- 50kHz). The corresponding dielectric loss are shown in the inset. (b) Temperature dependent capacitance measured at 30 kHz for the same sample under different hydrostatic pressure.

For each pressure value, the capacitance curve shows similar anomaly at $T_N$, marking the emerging of spin-induced ferroelectric phase. As pressure grows, $T_N$ increases gradually, consistent with corresponding susceptibility results. The data for P = 0.45 GPa are acquired by decreasing pressure (releasing process) from the maximal 0.83 GPa, the anomalies are reversibly observed (the blue curve), suggesting no degradation of the sample in high pressure medium.

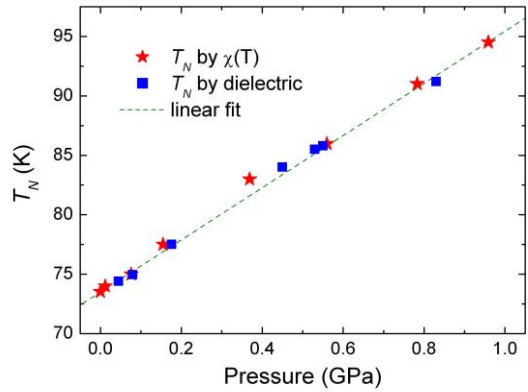

Fig. 4. The pressure dependent $T_N$ of CuBr$_2$. $T_N$ is determined by our magnetic (red) and dielectric (blue) measurements. The dashed line are linear fit.

In Fig. 5, we summarize all the $T_N$ values of $CuBr_2$ under different pressure, which is determined by the consistent magnetic susceptibility and dielectric measurements. $T_N$ increases almost linearly with growing pressure with the rate of $dT_N/dP \sim 21.9$ K/GPa. And no sign of saturation is observed up to 1 GPa (the limit of our apparatus). Till now, to our knowledge, such giant pressure sensitivity of the critical temperature has not been reported in other magnetic systems.

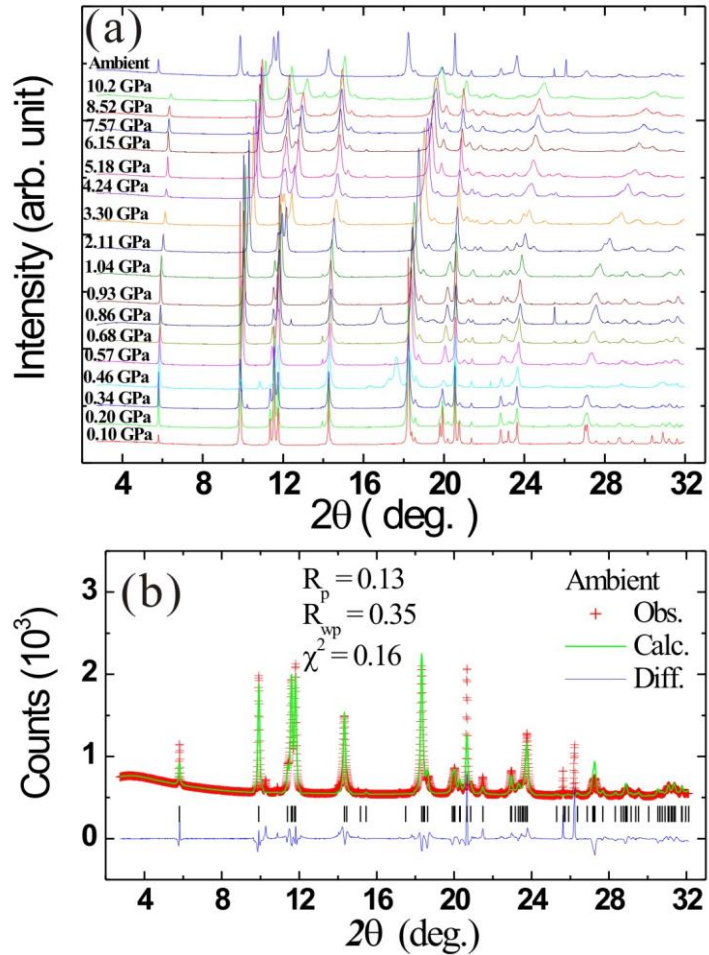

Fig. 5. (a) Powder x-ray diffraction profiles of $CuBr_2$ measured with DAC under different pressures. Curves are shifted vertically for clarity, (b) Rietveld refinement of the XRD pattern at ambient pressure. The observed (red), calculated (green) and the difference curves (blue) are plotted. The vertical black bars mark the positions of Bragg peaks. The refinement is carried out using GSAS software.

The structural stability of $CuBr_2$ under high pressure is studied via HP-XRD experiments at room temperature in the SPring-8 synchrotron. As shown in Fig. 5(a), upon loading process, P increases from 0.1GPa (red line at bottom) to the maximal 10.2GPa (green one near the top), the Bragg peaks shift gradually to higher angle due to the shrinkage of the $CuBr_2$ lattice. No splitting or emerging additional peaks were observed in our measurements. There are some impurity peaks appears at several pressures, e.g. the hump around 16.8° in the P = 0.86 GPa case, which comes from the gasket or holder materials used in the DAC device we used, due to the imperfect alignment upon sample mounting.

Near the end of our HP-XRD experiment, the pressure in DAC was released completely and the XRD for the P = 0 case (ambient) were measures to check posible degradation of the $CuBr_2$ sample. As shown as the blue profile at the topmost of Fig. 5(a), the pure monoclinic phase of $CuBr_2$ are confirmed, without any sign of the lattice collapse or decomposition by the loading process. Refinement with the GSAS package are carried out with each profile to acquire the accurate evolution of lattice parameters with pressure. There is no structural transition were observed, all the profiles are refined with the space group $C2/m$.

A typical refinement also shown in Fig. 4b, and the acquired lattice parameter is very close to the previous reports [26]. Due to highly layered structure of $CuBr_2$, there exists very strong preferred orientation in our powder sample. The corresponding refinement criteria, $R_p$ ( 13% ) and $\chi^2$ (0.16), which is slightly high.

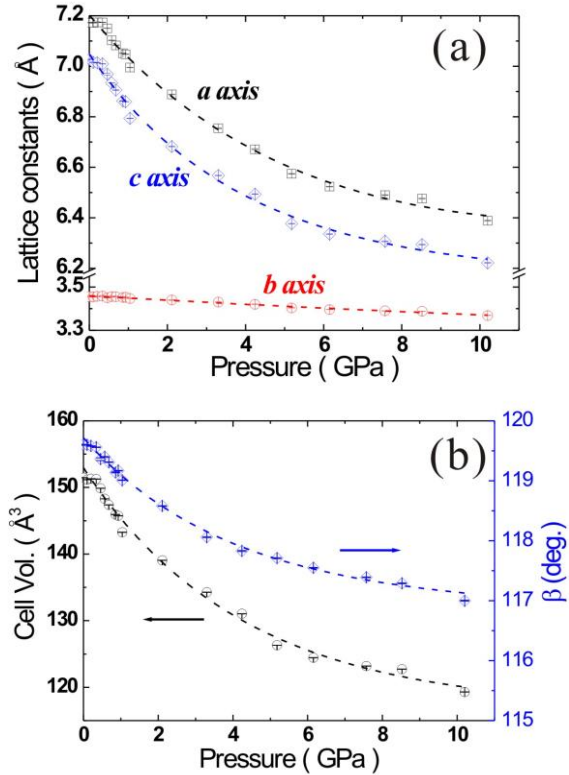

Fig. 6. Pressure dependence of lattice parameters. (a) *a*- *b*- and *c*-axis parameters. (b) β angle and cell volume.

The monoclinic lattice parameters from our refinements are summarize in Fig 6.   Upon increasing pressure, as shown in Fig.6 (a), the most remarkable feature is that the *a*- and *c*-axis lengths shrink much more strongly than *b*-axis in $CuBr_2$. From ambient to the highest pressure (10.2 GPa), the relative change of *a*- and *c*-axis reaches 10.9% and 11.5%, while *b*-axis just shrinks by only 2.5% and β angle changes slightly from 119.5 to 117 °. The spin chain remains very rigid. The relative decrease of cell volume reaches more than 21%, mainly coming from the great shrinkage of transverse direction, other than the slight contraction along the spin chain direction.

As has been noted, no pressure-induced additional phases is observed in $CuBr_2$. Multiferroicity emerges below $T_N$, which is induced by the spin spiral ordering propagating along b-axis below $T_N$.   According to the Anderson-Kanamori-Goodenough rule, the NN exchange interaction weakly ferromagnetic and NNN exchange interaction is antiferromagnetic. The competing between these two interactions results in the helical magnetic ordering in CuBr$_2$. But in low dimensional spin systems as CuBr$_2$, there exist strong fluctuations, impeding the emergence of long range ordering. The decrease of the dimensions along transverse directions (*a*- and *c*-axis) enhances the inter-chain coupling, hence suppressing the spin fluctuations. Meanwhile, the spin chain itself keeps rigid, change very slightly with pressure. Therefore, the intra-chain frustrated NN and NNN exchanges have little variation, which promote the long-range helical spin ordering in CuBr$_2$ at higher temperature, as well as the concomitant ferroelectric polarization. The large anisotropic compressibility of CuBr$_2$ lead to the enhanced critical temperature mainly from the pressure-enhanced inter-chain coupling. The similar effect is expected in other isostructural compounds as CuCl$_2$.

## 4. Summary

In summary, the hydrostatic pressure effect on the multiferroic quasi-1D spin-chain CuBr$_2$ have been studied via magnetic and dielectric measurements up to about 1 GPa. We find the giant enhancement of multiferroic critical temperature by exerting pressure (more than 20 K/GPa), while its dielectric dissipation keeps very low. The structural characterization via HP-XRD reveals that the highly anisotropic compressibility existing in CuBr$_2$. The external pressure compresses *a*- and *c*-axis dimensions much more effective rather than *b*-axis. The increasing of spin ordering temperature is considered to come from the enhanced the inter-chain coupling.

The multiferroic materials is naturally rare [27] and known candidates are still far from practical application. Our finding opens up a new potential way towards high temperature multiferroics. The critical temperature of known multiferroic compounds can be push to much higher temperature via tuning anisotropic exchange interactions by exerting pressure, chemical pressure, or appropriate strain. Especially there are many serendipities among the non-oxides systems may lead to great-leap-forward development to spin-driven multiferroic device application at room temperature.

Additional: During the preparation of this work, we notice that there is a preprints [28] on CuBr$_2$ reporting consistent conclusions with ours, extending the pressure range to up to around 4 GPa in which the linear pressure dependence of multiferroic critical temperature keeps well.

## Acknowledgments

The authors would like to thank Prof. Yang-Yuan Chen and Dr Ping-Chung Lee, Chi-Lian Hung and Min-Hsueh Wen for their great help with experimental details, and we are really indebted for the support from Academia Sinica and the National Science Council of Taiwan.